\documentclass[12pt]{article}
\usepackage[dvips]{graphics}
\usepackage{pdproc}
\usepackage{epsfig}

  \makeatletter 
  \def\@cite#1{[#1]} 
  \makeatother    
\newcommand{\be}{\begin{equation}} \newcommand{\ee}{\end{equation}}

\newcommand{\ba}{\begin{eqnarray}}
\newcommand{\ea}{\end{eqnarray}}

\newcommand{\bq}{\begin{equation}}
\newcommand{\eq}{\end{equation}}
\newcommand{\bqa}{\begin{eqnarray}}
\newcommand{\eqa}{\end{eqnarray}}
\newcommand{\ben}{\begin{enumerate}}
\newcommand{\een}{\end{enumerate}}
\newcommand{\bc}{\begin{center}}
\newcommand{\ec}{\end{center}}
\newcommand{\bqb}{\begin{eqnarray*}}
\newcommand{\eqb}{\end{eqnarray*}}

\newcommand{\gsim}{\;\raisebox{-0.9ex}
           {$\textstyle\stackrel{\textstyle >}{\sim}$}\;}
\newcommand{\lsim}{\;\raisebox{-0.9ex}{$\textstyle\stackrel{\textstyle<}
           {\sim}$}\;}

\newcommand\ch{{\tilde{\chi}}}

\newcommand{\PR}{\delta_{b R}}
\newcommand{\PL}{\delta_{b L}}

\newlength{\figwidth}
\newlength{\smallspace}
\newcommand{\vspa}{\vspace{\smallspace}}

\begin{document}

\renewcommand{\thefootnote}{\alph{footnote}}

\title{
Supersymmetry Tests from a Combined Analysis\\
 of Chargino, Neutralino, and Charged Higgs Boson\\
 Pair Production at a 1 TeV Linear Collider
}


\author{H. Eberl$^a$ with M. Beccaria$^{b,c}$, 
F.M. Renard$^d$, C. Verzegnassi$^{e, f}$ 
}


\address{
$^a$Institut f\"ur Hochenergiephysik der \"OAW, Vienna, Austria\\
$^b$Dipartimento di Fisica, Universit\`a di
Lecce, $^c$INFN, Sezione di Lecce, Italy\\
$^d$Physique Math\'{e}matique et Th\'{e}orique, UMR 5825,
Universit\'{e} Montpellier II, France\\
$^e$Dipartimento di Fisica Teorica, Universit\`a di Trieste, $^f$INFN, Sezione di Trieste, Italy}

\abstract{
We consider the production of chargino, neutralino and charged Higgs boson
pairs at future linear colliders for c. m. energies in the one TeV range
within the MSSM. We compute the leading (double) and next-to leading
(linear) supersymmetric logarithmic terms of the so-called "Sudakov
expansion" at one-loop level. We show that a combined analysis of the
slopes of the chargino, neutralino, and charged Higgs boson pair production
cross sections would offer a simple possibility for determining $\tan\beta$
for large ($\gsim$ 10) values, and the parameters $M_1$, $M_2$, and $\mu$. This
test could provide a strong consistency check of the considered
supersymmetric model.
}

\normalsize\baselineskip=15pt

\section{Introduction}

The Minimal Supersymmetric Standard Model (MSSM) 
looks still the most attractive extension of the Standard Model. 
In this model we have two charged bosons $H^{\pm}$,
two charginos $\ch^+_i$, mixtures of the W-inos and charged Higgsinos,
and four neutralinos $\ch^0_j$ which are mixtures of a B-ino, 
Z-ino and the two neutral Higgsinos. Realistic hope exists 
within the elementary particle physics community that at the
upcoming experiments at 
Tevatron~\cite{Tevatron} and LHC~\cite{LHC} 
will finally reveal the existence of supersymmetric particles, via direct
production of sparticle-antisparticle pairs. A logical and necessary next
step then is to measure the properties of the underlying theory to get access
to a more unified theory. For the MSSM case, a high luminosity
linear $e^+e^-$ collider~\cite{LC}
working in the TeV energy region will be the appropriate tool for that purpose.
This machine should be sufficiently accurate to 
provide the same kind of consistency tests of the model that were achieved in the
one and two hundred GeV region at LEP1 and LEP2 for the Standard Model, 
from detailed analyses of several
independent one-loop virtual effects.

In this contribution we show how to get access to some of the parameters of the
MSSM parameters. We will do a combined analysis 
of charged Higgs boson, chargino, and neutralino pair production at a 
linear $e^+e^-$ collider working at a CMS energy $\sqrt{s} \sim$~1~TeV.
The production cross sections have been calculated at one-loop level 
in the
\mbox{"Sudakov"} approximation \cite{Melles}.

\section{One-loop results}

If $\sqrt{s}$ is sufficiently larger than all the masses of the particles involved
in the process, a logarithmic Sudakov expansion can be adopted. At a 1 TeV collider
this is fullfilled for a SUSY mass scale $M_{\rm SUSY} \lsim 350$~GeV.  
The universal, process independent part has the form $(a \ln s/M_W^2 - ln^2 s/M_W^2)$.  
For the complete results for the production cross section 
$e^+e^- \to H^+ H^-,\, \ch_i^+ \ch_j^-,\,
\ch_k^0 \ch_l^0$ within this aprroximation we refer to 
\cite{e+e-chargedHiggses,e+e-charginos,e+e-neutralinos}.\\

The full amplitude $A$ can be decomposed into tree-level and one-loop part,\\
\mbox{$A = A^{\rm tree} + A^{\rm 1 loop}$}, 
with the one-loop part consisting of three parts, 
\begin{equation}
  A^{\rm 1 loop} = A^{\rm RG}  + A^{\rm non univ} + A^{\rm univ}\, .
\end{equation}
The RG (renormalization group) contribution stems from the
running behavior of the couplings. The non-universal,
process dependent contribution is the scatter angle $\vartheta$
dependent part of  
the box graphs and the third term is the $\vartheta$ independent 
universal contribution.\\[-0mm] 
 
The {\it RG contribution} $A^{\rm RG}$ gives linear logs
generated by the "running" of $g$ and $g'$
\begin{equation}
 A^{\rm RG}=-{1\over4\pi^2}~\left(g^4{\tilde \beta}_0{dA^{\rm tree}
\over dg^2}+
~g^{'4}{\tilde \beta}_0^\prime{dA^{\rm tree}
\over dg^{'2}}~\right)\  \log{s\over\mu^2}
\label{RGder}
\end{equation}
\noindent
$\mu$ is the scale where the numerical values
of $g$, and $g'$ are defined.
The $\beta$-functions in the MSSM 
are: ${\tilde \beta}_0 = -1/4$ and 
${\tilde \beta}_0^\prime = -11/4$.\\[-0mm]

The {\it non universal contribution} can be written as
$A^{\rm non ~univ} = A^{\rm tree} \cdot c^{\rm ang}$.
They are logarithmically dependent on $t/s =-\left( 1-\cos \vartheta \right)/2$
and $u/s=-\left( 1+\cos \vartheta \right)/2$.\\[-0mm]

The {\it universal contribution} $A^{\rm univ}$
can be of quadratic and of linear log type,\\ 
produced by initial vertex, and final vertex
and box diagrams,
$A^{\rm univ} = A^{\rm in} + A^{\rm fin}$.\\[0mm]

In all three productions the correction to the initial $e^+e^-$ lines
can be written in terms of tree-level structures with the 
coefficients
\begin{equation}
c^{\rm in}_{\alpha} = \frac{1}{16\pi^2}\left( g^2
I_{e^-_{\alpha}}(I_{e^-_{\alpha}}+1)+
  {g^\prime}^2\frac{Y^2_{e^-_{\alpha}}}{4} \right)
\left( 2 \log \frac{s}{M_W^2} -\log^2\frac{s}{M_W^2}\right)\, ,\quad
\alpha = L,R\, .
\label{cin}
\end{equation}

The universal term $A^{\rm fin}$ is different for $H^+ H^-$, $\ch_i^+ \ch_j^-$, 
and $\ch_k^0 \ch_l^0$ production.
In general, the total amplitude can be written as a combination 
of a S-, T-, and U-term, with the coefficients
\begin{equation}
A^{ab}_{ij} \equiv {e^2\over s}S^{ab}_{ij}+{e^2\over u}U^{ab}_{ij}
+{e^2\over t}T^{ab}_{ij}\, ,\quad a, b = L,R\, .
\end{equation}

{\it Charged Higgs boson pair production}: 
the total relative one-loop correction $\Delta(q^2)$ can be written as
\begin{equation}
\Delta(q^2)={\sigma^{{\rm tree +1 loop}}-\sigma^{{\rm tree}}\over\sigma^{{\rm tree}}} \, .
\end{equation}
In this case the one-loop Sudakov expansion is a very compact expression,
\begin{equation}
\Delta(q^2)= -~({3\alpha\over4\pi s^2_W M^2_W})(m^2_t\cot^2\beta+m^2_b \tan^2\beta)
\log{q^2\over M_W^2} + \Delta_{\rm RG}(q^2)\, , 
\end{equation}
dependent only on one SUSY parameter, on $\tan\beta$.\\

{\it Chargino pair production}: $A^{\rm fin}$ has many 
contribution. Only the Yukawa part is shown here, 
\begin{equation}
S^{\rm fin}_{ij} = {\alpha\over4\pi} \sum_{k=1}^2
S^{\rm tree}_{ik}.c_{kj}^{\rm Yuk} + \ldots \,, \qquad {\rm with}
\end{equation}
\begin{equation}
\label{Yukawa}
c_{kj}^{\rm Yuk} = -\frac{3\alpha}{2 s_W^2 M_W^2}
\bigg[
{m^2_t\over \sin^2\beta}
Z^{+*}_{2k}Z^{+}_{2j} \PL  +   
+{{m^2_b\over \cos^2\beta}}
Z^{-*}_{2k}Z^{-}_{2j} \PR \bigg]  \log{s\over M_W^2} \, .
\end{equation}
 
{\it Neutralino pair production}: $A^{\rm fin}$ has contributions from
all three possible structures, 
\begin{equation}
S^{\rm fin}_{ij} = {\alpha\over4\pi} \sum_{k=1}^4
S^{\rm tree}_{ik}  \left[c^{\rm fin~S~gauge}_{kj}
+ c^{\rm fin~Yuk}_{kj}\right]\,, ~~
X^{\rm fin}_{ij}={\alpha\over4\pi} \sum_{k=1}^4 X^{\rm tree}_{ik} \left[c^{\rm fin~gauge}_{kj}\right]\, ,
\end{equation}
with $X = U, T$, and symmetrized indizes $i$ and $j$.
The coefficients are
\begin{eqnarray*}
c^{\rm fin~S~gauge}_{kj} & = &  
{1+2c^2_{\rm w}\over 8 s^2_{\rm w}c^2_{\rm w}}
\bigg[
\left(Z^{N*}_{4k}Z^{N}_{4j}+ Z^{N*}_{3k}Z^{N}_{3j}\right) \PL  + ({\rm h.c.}) \PR \bigg]
\left(2\log{s\over M_W^2}-\log^2{s\over M_W^2}\right)\, , \nonumber\\
c^{\rm fin~gauge}_{kj}&=& -\frac{1}{s_{\rm w}^2}
\bigg[Z^{N*}_{2k}Z^N_{2j}\PL+Z^{N}_{2k}Z^{N*}_{2j}\PR \bigg] \log^2\frac{s}{M_W^2}\, ,\nonumber\\
c^{\rm fin~Yuk}_{kj}& = &
-{3\over 4 s^2_{\rm w} M^2_{\rm w}}
\bigg[\left({{m^2_t\over \sin^2\beta}} Z^{N*}_{4k}Z^{N}_{4j} +
{{m^2_b\over \cos^2\beta}} Z^{N*}_{3k}Z^{N}_{3j}\right) P_L  + ({\rm h.c.}) P_R \bigg]  
\log{s\over M_W^2}\, .
\end{eqnarray*}
The tree level coefficients can be found in \cite{e+e-charginos,e+e-neutralinos}. 
Note, that the notation $Z^+ \equiv V$, $Z^- \equiv U$, $Z^N \equiv Z$,
and e.g. $S^{\rm tree}_{ik} c^{\rm fin~Yuk}_{kj} \equiv S^{a b,\, \rm tree}_{ik} c^{\rm fin~Yuk}_{b,\, kj}$ is used. 

\section{Numerical results}

We show results for three scenarios, $S_1$, $S_2$, and $S_3$. The scenario $S_1$ 
is the Tesla benchmark point RR2\cite{RR2} with the two lightest neutralinos being respectively 95\% bino 
and 82\% wino. The set $S_2$ is a mixed scenario with neutralinos having non negligible
gaugino and Higgsino components; $\ch^0_1$ is 86\% bino and 13\% Higgsino, 
$\ch^0_2$ is 11\% bino, 48\% wino and 41\% Higgsino. 
Finally, $S_3$ is a purely Higgsino one  with the 
two lightest neutralinos being 92\% and 98\% Higgsino like.
The values of the input parameters as well as the masses of the two charginos and of 
the two lightest neutralinos are summarized in Tab.~(1).
We have performed a standard $\chi^2$ analysis assuming in the various 
scenarios 10-12 experimental points at energies ranging from 700-850 GeV 
(depending on the scenario) up to 1200 GeV and assuming an experimental accuracy of 
1\% for the cross sections of chargino, and neutralino, and 2\% for that of charged Higgs 
boson pair production.\\
\begin{table}
\caption{\label{tab:table1}Input parameters and masses of charginos, and lightest neutralinos 
for the three input sets $S_1$, $S_2$, and $S_3$. All parameter are given in GeV, 
and $\tan\beta$ = 30 in all three scenarios.}
\begin{center}
\begin{tabular}{c|ccc|cccc|}
      & $M_1$ & $M_2$ & $\mu$ & $\ch^\pm_1$ & $\ch^\pm_2$ & $\ch^0_1$ & $\ch^0_2$ \\
\hline
$S_1$ & 78  &  150 & 263 & 132 &  295 &  75 & 133 \\
$S_2$ & 100 &  200 & 200 & 149 &  266 &  92 & 153 \\
$S_3$ & 200 &  400 & 100 &  95 &  417 &  82 & 109
\end{tabular}
\end{center}
\end{table}
\begin{figure}[h!]
\begin{center}
\vspa
\mbox{\resizebox{\figwidth}{!}{\includegraphics{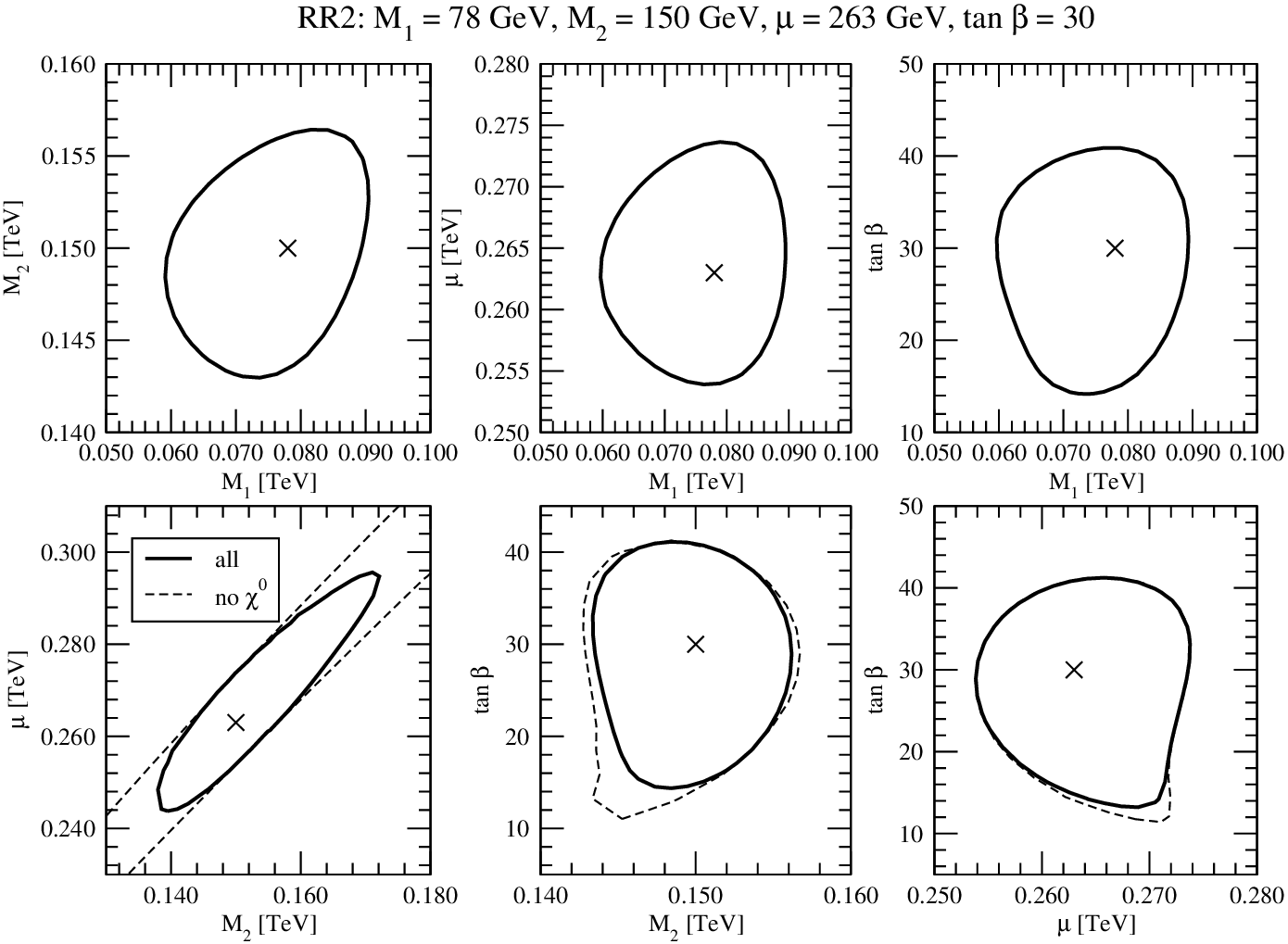}}} 
\vspace{-3mm}
\caption{$S_1$ scenario, $1\sigma$ error bounds on the MSSM parameters $M_1$, $M_2$, $\mu$, and $\tan\beta$.
In this and in the following figures the crosses denote the values of the parameters in the 
specific scenario.}
\label{fig:RR2}
\end{center}
\end{figure}
\noindent
\begin{figure}[h!]
\begin{center}
\vspa
\mbox{\resizebox{\figwidth}{!}{\includegraphics{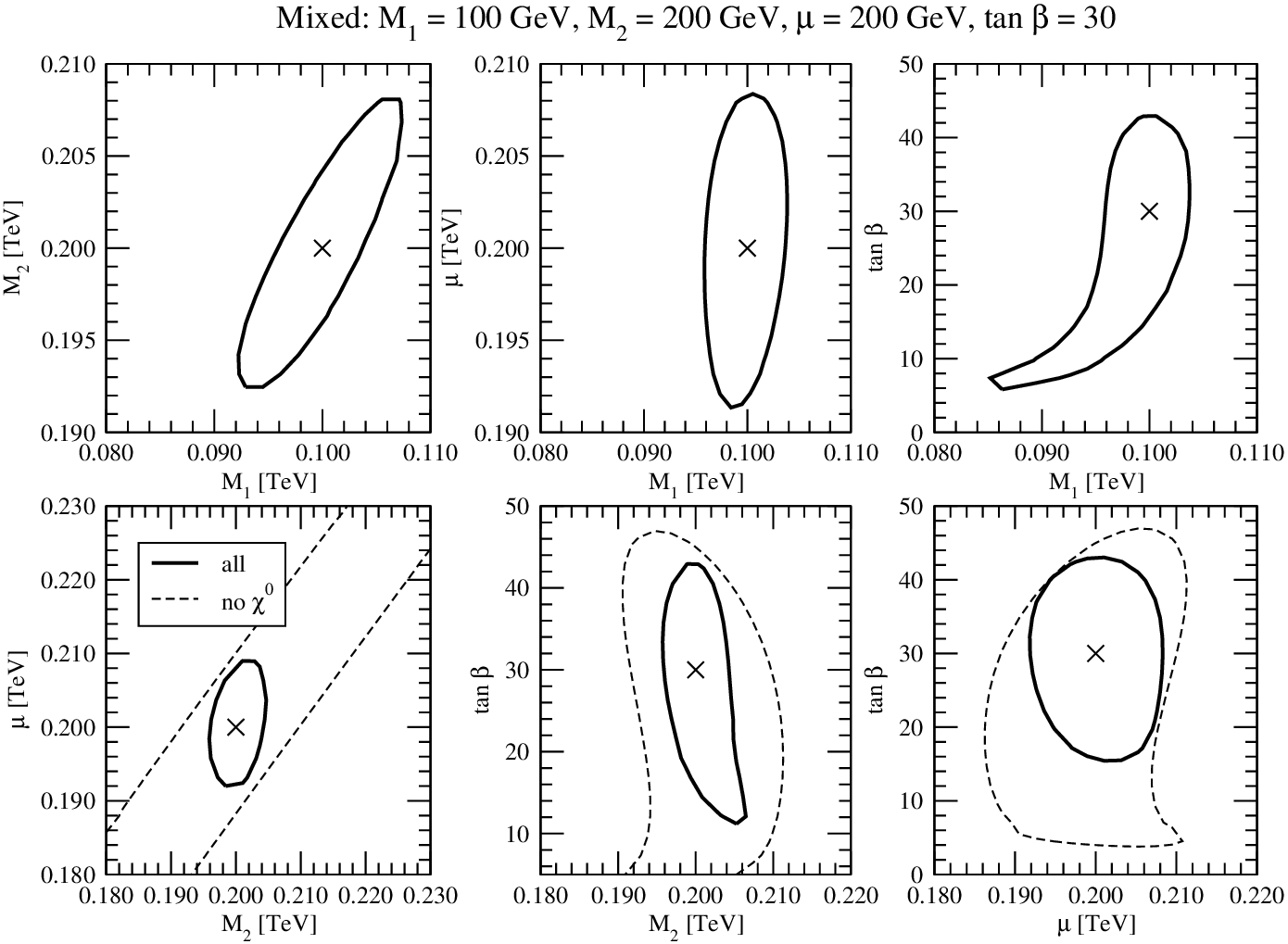}}}  
\vspace{-3mm}
\caption{$S_2$ scenario, $1\sigma$ error bounds on the MSSM parameters $M_1$, $M_2$, $\mu$, and $\tan\beta$.}
\label{fig:SPS4}
\end{center}
\end{figure}
\noindent
\begin{figure}[h!]
\begin{center}
\vspa
\mbox{\resizebox{\figwidth}{!}{\includegraphics{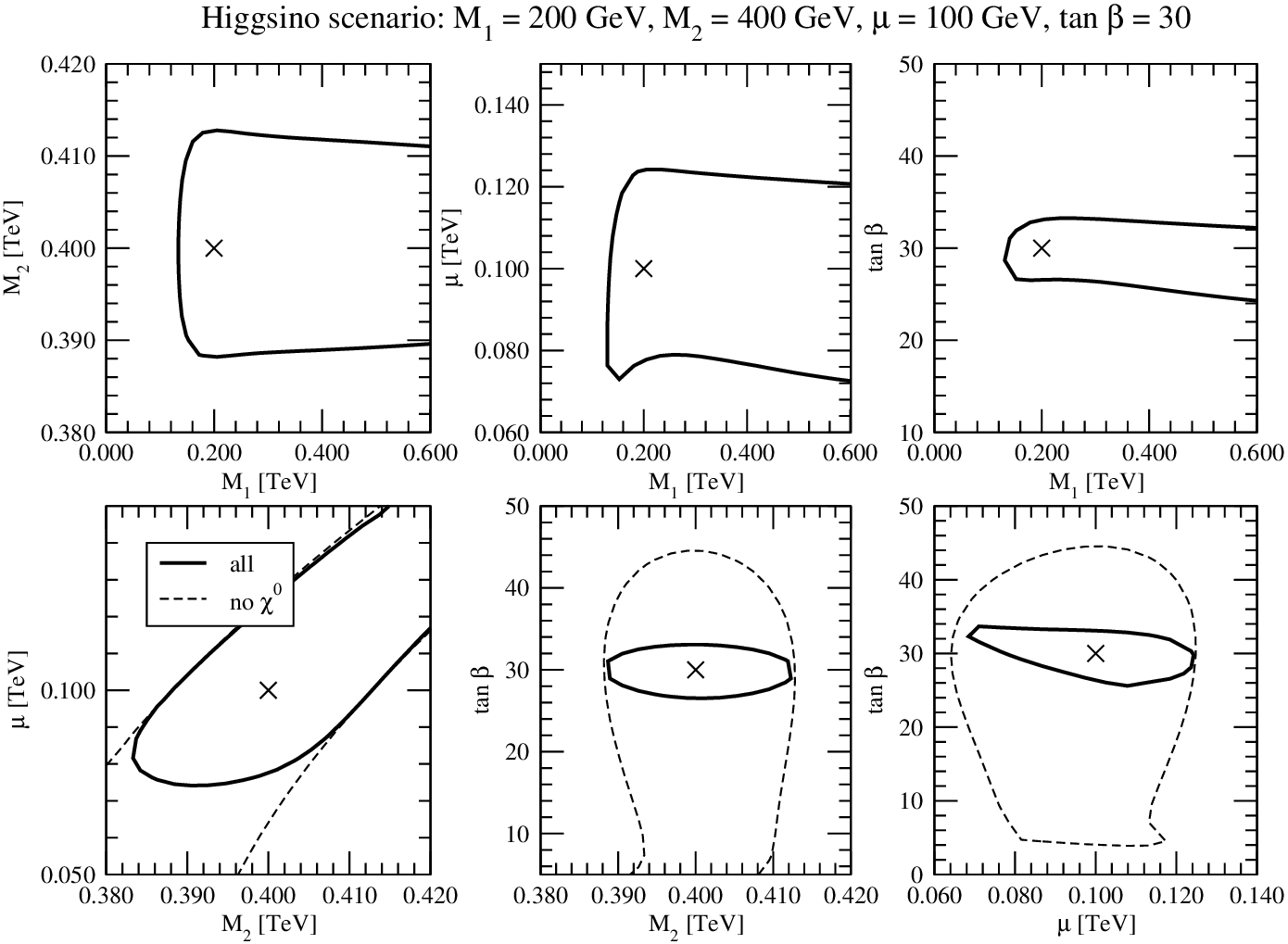}}} 
\vspace{-3mm}
\caption{$S_3$ scenario, 
$1\sigma$ error bounds on the MSSM parameters $M_1$, $M_2$, $\mu$, and $\tan\beta$.}
\label{fig:Higgsino}
\end{center}
\end{figure}
\noindent
All three figures have the same order of contour plots.
The upper ones are the planes $(M_1, M_2)$, $(M_1, \mu)$, $(M_1, \tan\beta)$
and the lower ones the planes $(M_2, \mu)$, $(M_2, \tan\beta)$, and $(\mu, \tan\beta)$.
The dashed lines denote the results based on the analysis in \cite{e+e-charginos}, where the 
neutralino channels are not yet included. In Fig.~1 we see that a combined analysis
gives closed areas for all six planes, also for $(M_2, \mu)$, but the bounds on
$(M_2, \tan\beta)$, and $(\mu, \tan\beta)$ practically do not improve.
The results for $S_2$ are similar, but the improvement is better now, especially
in the $(M_2, \mu)$ plane. In the Higgsino scenario $S_3$ the two light neutralinos
are quite insensitive to the Higgsino component. Therefore there is no upper bound
on $M_1$, as can be seen in the first three figures. The error on $\tan\beta$ is 
reduced from $\sim$ 30-40\% in the first two figures to $\sim$ 10\%.   

From a mass measurement only, one cannot deduce 
a 1-1 correspondence between measured masses and the 
parameters $M_1$, $M_2$, $\mu$, and $\tan\beta$. Different
sets of parameters can 
reproduce ``essentially'' the same masses~\cite{MassesVsParameters}.
The $\chi^2$ analysis presented in this work helps to disentangle
this ambiguity. 

\bibliographystyle{plain}

\end{document}